# Implementing Snort Intrusion Prevention System (IPS) for Network Forensic Analysis

Kashif Ishaq[1], Hafiz Ahsan Javed[1]

[1]School of Systems and Technology, University of the Management and Technology, Lahore, Pakistan

**Corresponding Author:** kashif.ishaq@umt.edu.pk

**Abstract**

The security trade confidentiality, integrity and availability are the main pillar of the information systems as every organization emphasize of the security. From last few decades, digital data is the main asset for every digital or non-digital organization. The proliferation of easily accessible attack software on the internet has lowered the barrier for individuals without hacking skills to engage in malicious activities. An Industrial organization operates a server that (Confluence) serves as a learning platform for newly hired employees or Management training officers, thereby making it vulnerable to potential attacks using readily available internet-based software. To mitigate this risk, it is essential to implement a security system capable of detecting and preventing attacks, as well as conducting investigations. This research project aims to develop a comprehensive security system that can detect attack attempts, initiate preventive measures, and carry out investigations by analyzing attack logs. The study adopted a survey methodology and spanned a period of four months, from March 1, 2023, to June 31, 2023. The outcome of this research is a robust security system that effectively identifies attack attempts, blocks the attacker's IP address, and employs network forensic techniques for investigation purposes. The findings indicate that deploying Snort in IPS mode on PfSense enables the detection of attacks targeting e-learning servers, triggering automatic preventive measures such as IP address blocking. The alerts generated by Snort facilitate investigative actions through network forensics, allowing for accurate reporting on the detrimental effects of the attacks.

## I. Introduction

The rapid pace of technological advancement necessitates an increased focus on enhancing network security. This is particularly crucial given the growing availability of hacking and cracking knowledge, coupled with the easy accessibility of tools that facilitate such activities. These tools range from commonly used network utilities to specialized tools employed for carrying out malicious attacks [1].

There are three main pillar of security.

- Confidentiality
- Integrity
- Availability

**Confidentiality**

The first principle is Privacy, which emphasizes restricted access to information, allowing only authorized individuals to view it while preventing unauthorized individuals from accessing it. Additionally, even if unauthorized individuals do manage to gain access,[2] the information should be rendered insignificant or inconsequential to them. As an example, only I, as the authorized individual, have the privilege to access my online banking account [3].

**Integrity**

The next principle is Information Integrity[4], which safeguards the accuracy and consistency of data. It ensures that the information remains reliable and unchanged both within the dataset itself and within the organization or system hosting the data. Occasionally, inconsistencies may arise due to flaws in the system, but maintaining the integrity of data is crucial to its security. In essence, integrity guarantees that data remains in its original state, maintaining its validity, completeness, and structure. It is a vital aspect of data security [5].

**Availability**

The third principle is Validation, which verifies the authenticity of an individual's claimed privileges or access rights to information. It ensures that only those who genuinely possess the required permissions can gain entry. For instance, when attempting to access my online banking account, the validation process confirms my identity to prevent unauthorized access by someone else. Both data authenticity and availability play significant roles in data management. Availability ensures that data is accessible to authorized users, but only with the appropriate rights. Not everyone has administrative rights or unrestricted access to data, emphasizing the importance of proper validation and access control mechanisms [6].

Network forensics serves as a valuable approach in ensuring network security. When a computer network falls victim to an attack, conducting a thorough investigation becomes imperative. This investigation aims to uncover and gather relevant evidence associated with the attack[7]. The Organization server is highly susceptible to potential attacks due to the presence of critical data and its frequent usage by both Employees and MTO's. Neglecting to appropriately address and mitigate such attacks could significantly disrupt the learning activities [8].

Following an attack, it becomes crucial to identify the nature of the attack to optimize the network infrastructure. To achieve this, the implementation of a network-based Intrusion Prevention System (NIPS) is essential. By utilizing Snort integrated into PfSense, this security system can effectively block and investigate alerts for the purpose of network forensics [9].

## II. Literature Review

There is critical recap of what has already been researched on the subjected topic as well as it is been taken from books, journals, research papers etc. As per research information security is the most important part of the every network infrastructure and every organization wants to improve security infrastructure along with secure the data.

As per foundation knowledge of subjected topic and has been discovered so far the company is currently grappling with a significant challenge pertaining to the security of their confluence servers. These servers hold critical data related to essential learning materials and a wide array of notable learning activities. The increasing proliferation of malicious software across the internet has lowered the barriers of entry for individuals lacking specialized expertise, thereby enabling them to engage in nefarious activities. This unfortunate vulnerability exposes the servers to considerable risks within the overarching framework of the company's operations. A promising avenue to rectify this vulnerability lies in the implementation of a robust network security system, leveraging the capabilities of Snort tools. This sophisticated toolkit would empower the organization to proactively detect any anomalous activities transpiring across the network. To fortify the preventive measures, the deployment of Snort tools onto the PfSense open-source firewall emerges as a strategic approach. This integration would facilitate the swift identification and subsequent blocking of IP addresses linked to potential attackers.

Furthermore the purposed is the new gap and unresolved problem as well as to have the without duplicate of work and to avoid the conflicts with the previous studies and solution providing without duplicate work.

Moreover, in order to mount a comprehensive defense against potential attacks, it is prudent to adopt investigative measures. This can be achieved through the adept utilization of network forensic techniques. These techniques would play a pivotal role in uncovering the origins and methods of attacks, enabling the organization to not only neutralize ongoing threats but also bolster their future preparedness against sophisticated security breaches.

It is imperative for the company to address this pressing security concern with a multi-faceted approach, amalgamating proactive detection, responsive mitigation, and insightful investigation. By doing so, the company can fortify its defenses and ensure the utmost security of its invaluable data assets.

**There are features analysis that every firewall supports according to the information security.**

| Feature / Aspect | PFSENSE | SOPHOS | Fortinet | Panda Firewall |
|---|---|---|---|---|
| Firewall Type | Open Source | Commercial | Commercial | Commercial |
| User Interface | Web-based | Web-based, Central Management | Web-based, Central Management | Web-based, Central Management |
| Security Modules | Firewall, VPN, IDS/IPS, Content Filtering, Antivirus | Firewall, VPN, IDS/IPS, Web Filtering, Antivirus | Firewall, VPN, IDS/IPS, Web Filtering, Antivirus | Firewall, VPN, IDS/IPS, Web Filtering, Antivirus |
| Scalability | Good for Small to Medium Networks | Suitable for Small to Enterprise Networks | Suitable for Small to Enterprise Networks | Suitable for Small to Medium Networks |
| Performance | Dependent on Hardware Specs | Balanced Performance | Balanced Performance | Balanced Performance |
| Advanced Threat Protection | Limited | Included in Some Versions | Included in Some Versions | Included in Some Versions |
| Centralized Management | Not Included | Centralized Management Platform | Centralized Management Platform | Centralized Management Platform |
| Intrusion Detection/Prevention | Yes | Yes | Yes | Yes |
| Web Filtering | Limited | Yes | Yes | Yes |
| Application Control | Limited | Yes | Yes | Yes |
| VPN | Yes | Yes | Yes | Yes |
| Cloud Integration | Limited | Yes | Yes | Limited |
| Reporting | Basic | Detailed Reporting | Detailed Reporting | Detailed Reporting |
| Support | Community-based | Commercial Support Available | Commercial Support Available | Commercial Support Available |

**Comparison analysis with different firewalls**

Based on check at different firewalls below are the results of purposed research.

| Analysis |
|---|

| S# | Description | PFSENSE | Fortinet | Panda | Sophos |
|---|---|---|---|---|---|
| 1 | Packet capture | YES | YES | YES | YES |
| 2 | Packet filtering | YES | YES | YES | YES |
| 3 | Snort Package availability | YES | NO | NO | YES |
| 4 | Snort Package installation | YES | NO | NO | YES |
| 5 | Log Analyzer | YES | YES | YES | YES |
| 6 | IDS scanning | YES | YES | YES | YES |
| 7 | IPS scanning and action | YES | YES | YES | YES |

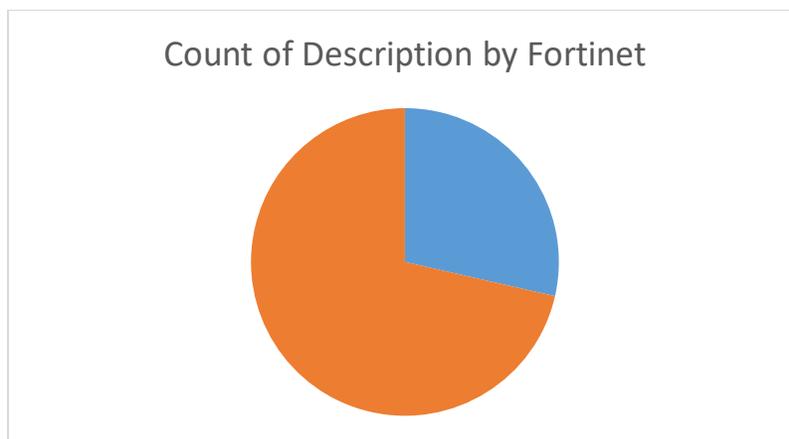

Count of Description by Fortinet

### III. Proposed Methodology

The challenge faced by the company revolves around the vulnerability of their confluence servers, which house critical data related to learning materials and various noticeable and learning activities. The increasing availability of attack software on the internet has made it possible for individuals without specialized expertise to engage in malicious activities. Consequently, this vulnerability puts the servers at risk within the context of the company[10].

One potential solution to address this issue is the implementation of a robust network security system utilizing Snort tools. These tools would enable the detection of any suspicious activities transpiring on the network. To enhance the preventive measures, Snort tools can be installed on the PfSense open-source firewall, allowing for the blocking of IP addresses associated with attackers. Furthermore, to effectively counter attacks, investigative measures can be undertaken by employing network forensic techniques[11].

By adopting this alternative solution, the company can enhance the security of their confluence servers, mitigate potential attacks, and safeguard the integrity of their learning activities[12].

TABLE 1 FORENSIC QUESTIONS SUMMARY

| Question for Forensic | THE ANSWER |
|---|---|

| RQ1: What were the high-quality publication channels for forensic research, and which geographical areas have been targeting forensic research over the years? | The primary aim of Research Question 1 (RQ1) was to systematically search for high-quality research articles in the field of information security and network system through reputable and widely recognized publication channels. Additionally, the study involved conducting a thorough quality assessment of the selected articles and extracting relevant meta-information. This process helped gather useful statistical data, such as geographical distribution and publication trends over the years. The objective was to ensure a comprehensive and reliable overview of the state of research in terms of both content quality and geographical coverage over time. |
|---|---|
| RQ2: What were the widely used theories, models, and frameworks proposed or adopted for forensic research? | In the realm of research, numerous theories, frameworks, and models are utilized to investigate various phenomena and address specific challenges. These methodologies play a crucial role in understanding complex systems and phenomena across different application domains. They provide a structured approach to analyze data, draw meaningful insights, and make informed decisions. In the field of computer networks and AI, researchers often leverage various theories and models to improve automation and network management. In the context of security systems for computer networks, researchers use machine learning models, such as Support Vector Machines (SVM) or Random Forests, to classify network traffic and detect anomalies. Bayesian networks are also employed to model and reason about network events and potential security threats, providing valuable insights for network administrators. |
| RQ3: What were different application domains for the snort forensic application, and in which various forms were these applications exposed for the end-users? | From a learning perspective, various target application areas in the security domain include network intrusion detection, malware classification, and anomaly detection. Different modes of exposition for these firewall applications involve rule-based systems, machine learning-based approaches, and hybrid models that combine both techniques for enhanced threat detection and prevention. |
| RQ4: What was the specific content adopted for teaching and learning in research? | To cater to diverse content needs for Snort applications, a comprehensive approach can be adopted, focusing on theoretical comprehension, hands-on practical exercises, and security perspectives. By offering a balanced blend of reading materials, practical lab sessions, and discussions on network infrastructure security, learners can gain a deeper understanding of Snort's usage, configuration, and its role in enhancing overall network protection. |

RQ5: How and in what different perspectives the security applications were evaluated, and what were the evaluation measures and tools used for their evaluation?

RQ6: Compare the usage of simple applications with snort security applications for learning?

By employing these evaluation measures, security professionals can gauge Snort's performance and suitability for network defense, leading to effective threat detection and improved network infrastructure security.

Conduct a comparative analysis to assess the efficacy of a basic security application for improved insights and performance evaluation.

RQ1: What were the high-quality publication channels for forensic research, and which geographical areas have been targeting forensic research over the years?

The primary objective of Research Question 1 (RQ1) was to perform a systematic search for well-regarded research articles in the realm of information security and network systems, published through reputable and widely recognized channels. Additionally, the study entailed conducting a rigorous assessment of the selected articles' quality and extracting pertinent meta-information. This meticulous process facilitated the acquisition of valuable statistical data, including insights into geographical distribution and publication trends spanning multiple years. The overarching aim was to present a comprehensive and dependable overview of the current state of research, encompassing both the quality of content and its geographic representation over time.

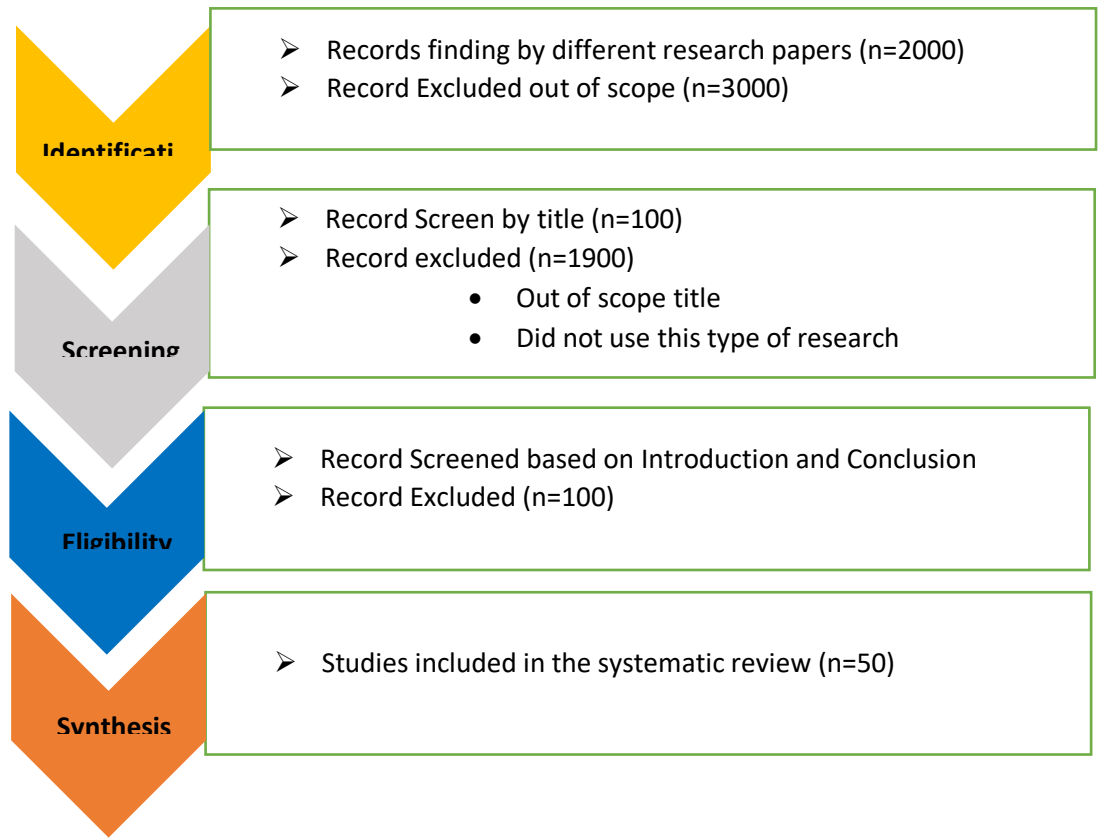

Selection of relevant articles using Systematic Review Process.

## IV. RESULT AND DISCUSSION

The network infrastructure of the company, although complex, exhibits vulnerabilities as some servers are accessible from external sources. The potential for attacks arising from these vulnerabilities poses a significant threat to the network's integrity[13]. To address this concern, the authors propose the integration of a forensic network server into the existing infrastructure, leveraging the PfSense open-source firewall and Snort tools. This addition aims to facilitate comprehensive analysis and investigation of attack origins, while also enabling the implementation of preventive measures, such as IP address blocking against identified attackers[14].

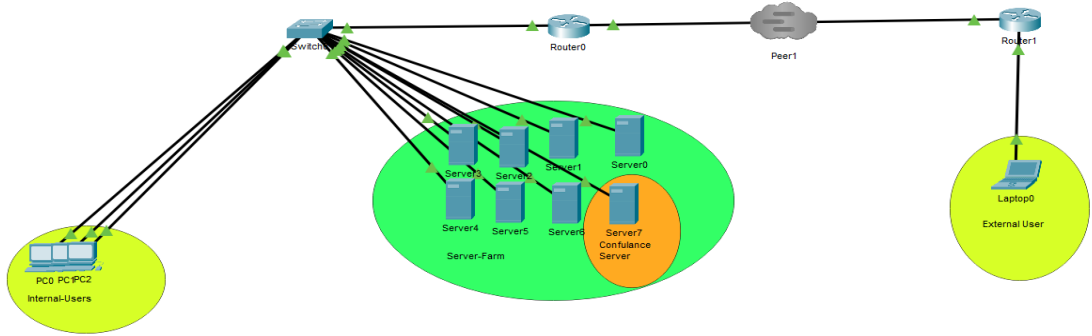

Fig.1. Proposed Network Topology

During the preliminary assessment[15], a Denial of Service (DoS) attack was conducted using the Low Orbit Ion Cannon (LOIC) tool while Snort, a network intrusion detection system installed on PfSense, remained inactive. The outcome of this test is as follows: [Here you can provide the results obtained without revealing specific details about the attack or any unethical actions].

- The below picture provides the result of the PfSense as per picture there are normal activities going on the network in the WAN and LAN[16].

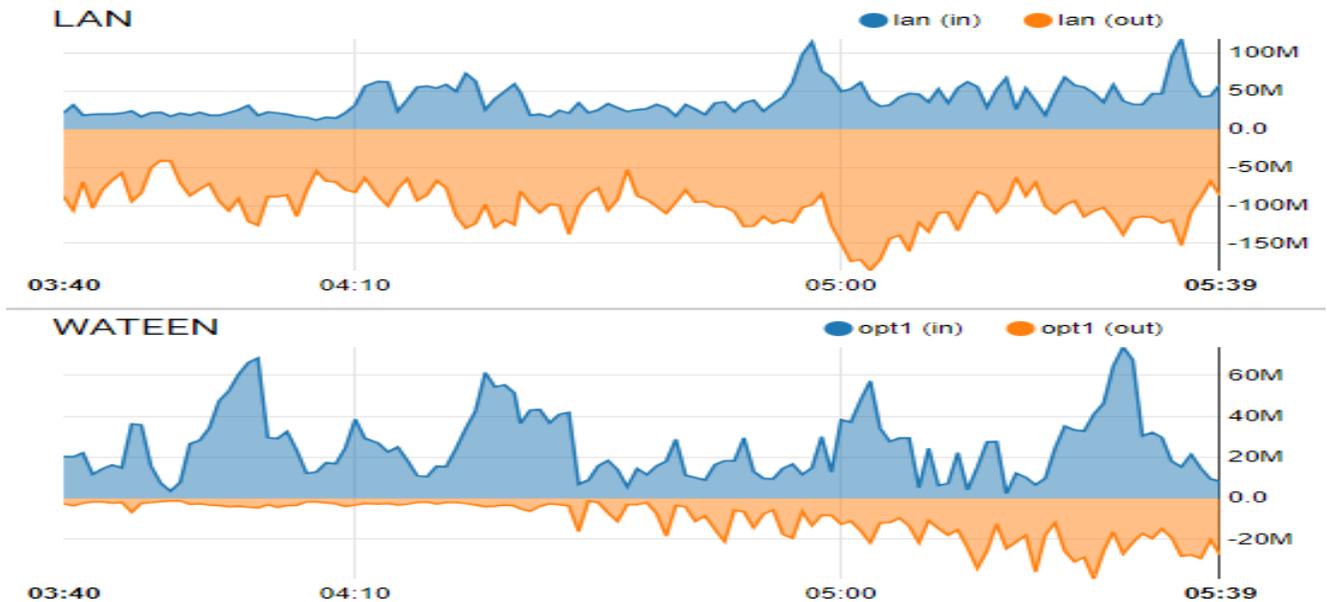

Fig.2. Network Traffic before Active Snort

- There are no any alerts appears on the firewall dashboard or any snort menu as well.

In the final phase of the assessment, Snort, which was previously installed on PfSense, was activated and utilized as a forensic server[17]. Various attack experiments were conducted during this phase. The focus was on analyzing the information gathered from the Alert tab and the Blocked tab within Snort for network forensic purposes[18]. These logs and data provided valuable insights into potential security breaches, aiding in the investigation and understanding of the attack scenarios for further analysis and improvements in the network's security posture[19].

1) HTTP Inspection

    Testing using HTTP inspection was conducted, wherein client computers attempted to exploit the forensic server websites. During this test, Snort effectively detected the HTTP inspection and generated alerts accordingly. In response to the detected attack, PfSense automatically took action by blocking the IP address of the attacker[20]. This successful detection and proactive blocking demonstrate the effectiveness of Snort and PfSense in mitigating potential threats and enhancing the security posture of the network[21].

Fig.3. HTTP Inspect Alert

The network forensic processes were executed following the forensic process model.

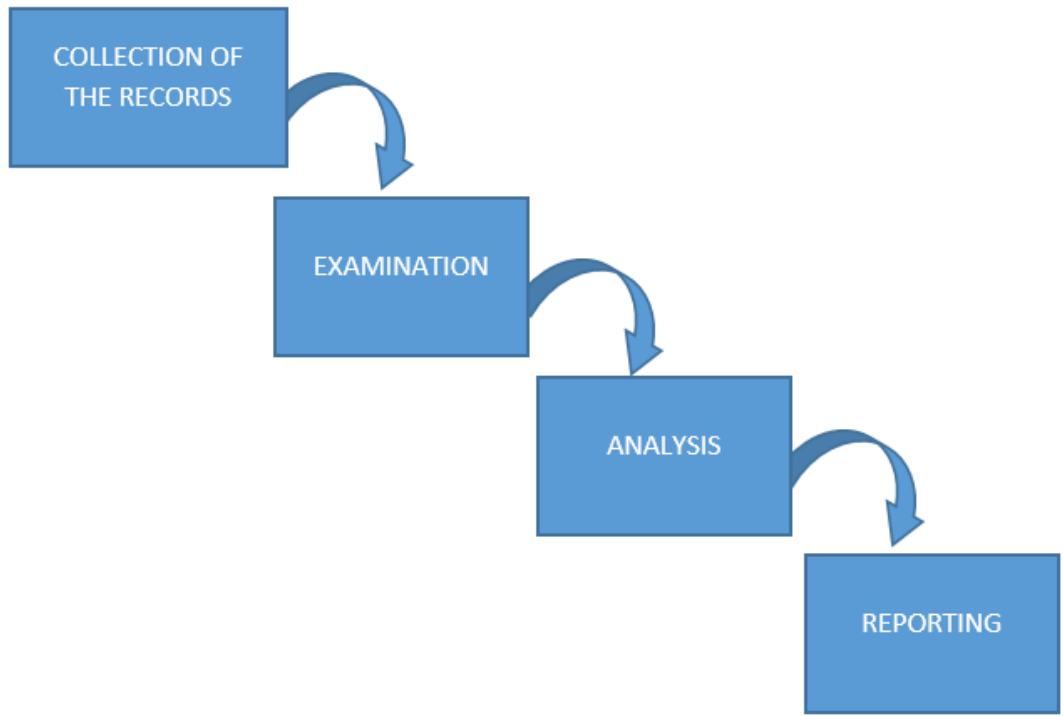

Fig.4. Forensic Process Model Approach

The analysis of alerts generated by Snort can be categorized into three stages:

1. Collection Stage: This initial stage involves the gathering of evidence triggered by Snort's detection of an attack, particularly in the form of HTTP Inspect. The information collected at this stage forms the basis for further investigation[22].
2. Examination Stage: At this stage, a comprehensive examination of the collected evidence takes place. The focus is on scrutinizing the generated alerts and any downloaded file alerts. Thorough checks are performed to ascertain the nature and scope of the detected attack[23].
3. Analysis Phase: In this critical stage, the results obtained from the examination are thoroughly studied and analyzed. The primary objective is to address specific forensic questions, gain insights into the attack's characteristics, and draw meaningful conclusions based on the evidence collected and examined[24].

Testing:

TABLE 1 FORENSIC QUESTIONS SUMMARY

| Question for Forensic | THE ANSWER |
| --- | --- |
| What specific attack that occurred? | Hypertext transfer protocol (HTTP) |
| When the attack occur? | 08-04-2023 jam 10.21, 07-04-2023 jam 20.48 |
| The IP Address of the attacker? | 148.229.33.150, 63.17.125.15 |
| The destination of the IP Address? | 10.0.5.188 (IP Address of the server) |
| The protocol is used? | Transmission control protocol (TCP) |

| How many ports were attacked? | The port was 80 |
|---|---|

- The reporting stage involves the documentation and creation of a comprehensive report detailing the entire inspection process and the information
  Obtained from the preceding stages. The report aims to provide a clear and structured account of the investigative procedures, the detected attack, the examination results, and the analysis findings, presenting a cohesive narrative that facilitates understanding and supports informed decision-making[25].
- A Denial of Service (DoS) attack was executed with the intention of disabling the confulance server, which holds the IP Address 10.0.5.188. The DoS attack originated from a client PC running the Debian Server operating system. The objective of this attack was to disrupt the normal functioning of the e-learning server by overwhelming it with a high volume of traffic or malicious requests, rendering it inaccessible to legitimate users[26].

Fig.5. Denial of Services

As alerts created by log of snort and used for held the forensic network by forensic questions.

Fig.6. Alert Denial of services

TABLE 2 SUMMARY OF FORENSIC QUESTIONS (2)

| Question for Forensic | THE ANSWER |
|---|---|
| What attack happened? | Hypertext transfer (HTTP) |
| When the attack occur? | ICMP |
| The IP Address of the attacker? | 148.229.33.150, 63.17.125.15 |
| The destination of the IP Address? | 10.0.5.188 (IP Address of the server) |

Port scans are conducted using Zen map software installed on the client's computer to identify potential vulnerabilities on the forensic servers. The purpose of these scans is to detect any open ports that may be susceptible to exploitation in order to assess the security posture of the servers[27].

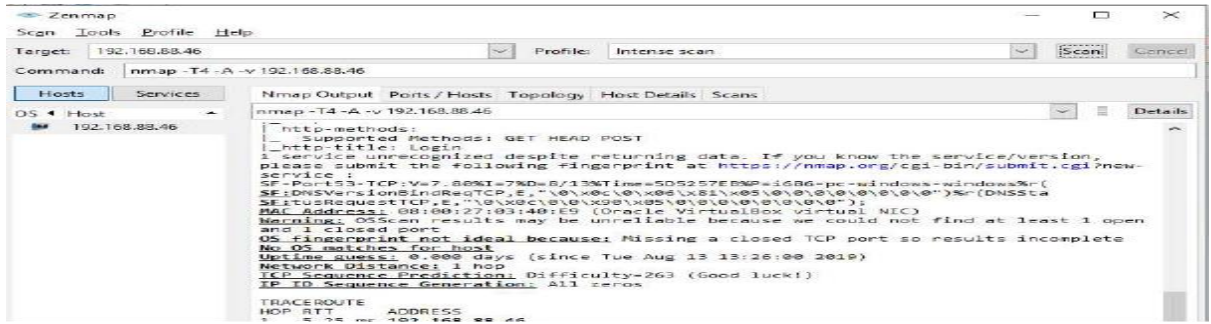

Fig.7. Port Scan

As alerts generated from the snort and these generated log can be used for analysis.

Fig.8. Alert Port Scan

TABLE 3 SUMMARY OF FORENSIC QUESTIONS (3)

| Question for Forensic | THE ANSWER |
|---|---|
| What attack happened? | 13-08 Jam 13.22 |
| When the attack occur? | ICMP |
| The IP Address of the attacker? | 200.229.33.150, 202.17.125.15 |
| The destination of the IP Address? | 10.0.5.188 (IP Address of the server) |

ARP spoofing is a technique employed to intercept data packets destined for victim PCs by manipulating the Address Resolution Protocol (ARP) tables. Snort, in this context, plays a crucial role as it offers rules preprocess, enabling[28]

The detection of ARP spoofing activities. The ARP spoofing is conducted from a PC client with the Ubuntu operating system, utilizing the Ettercap application for executing the test[29].

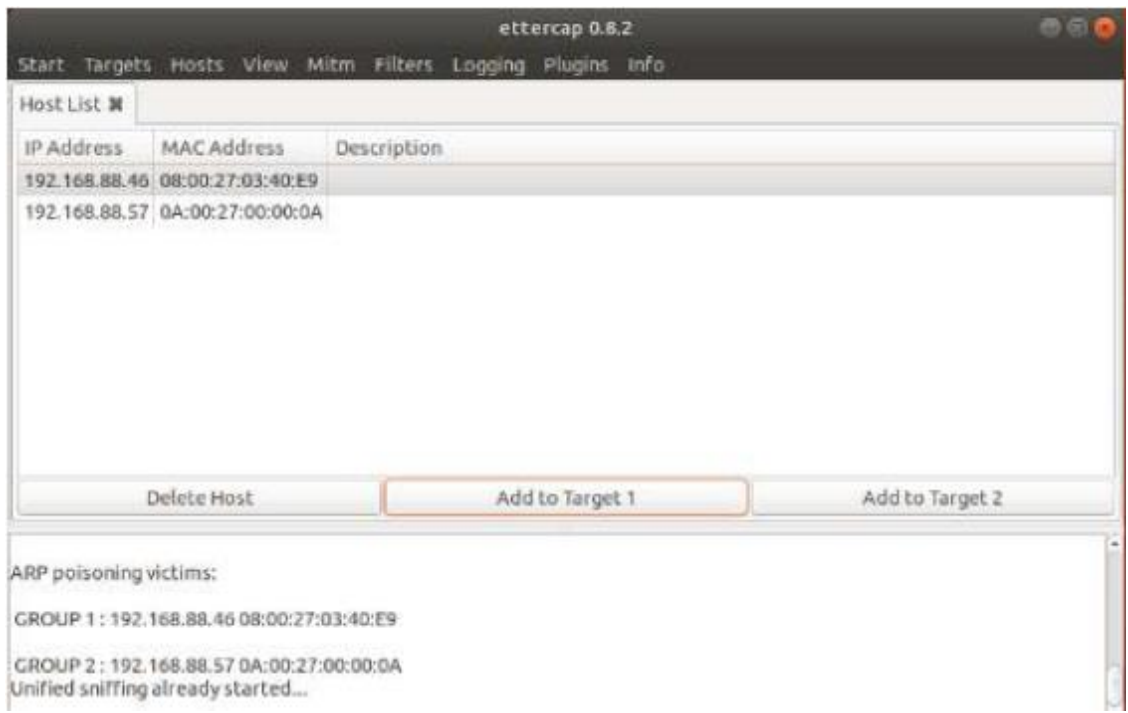

Fig.9. ARP Spoofing

The generated alerts from snort does not have the identification just because of the available firewall does not include the hacker/attacker identification[30].

Fig.10. Alert ARP Spoofing

Fig.11. Rules ARP Spoofing

Offending traffic. This proactive approach helps safeguard the network infrastructure by swiftly mitigating potential threats and enhancing overall security posture[30]

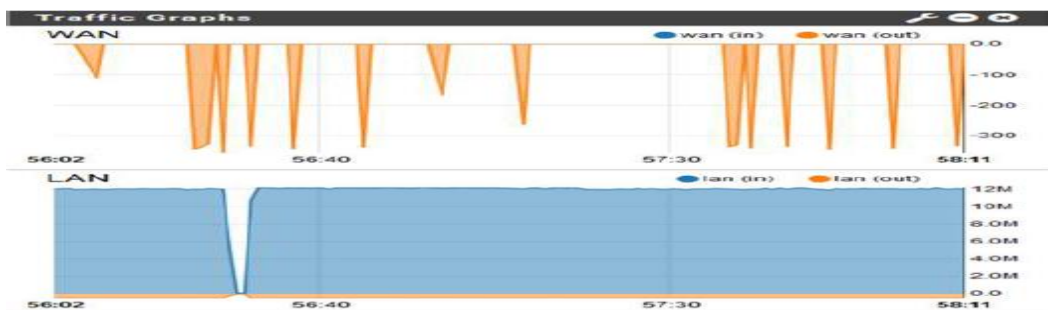
Fig.12. Log Blocked Snort

The results of the Snort test indicate that the alerts generated by the Snort log effectively detected each data packet involved in attack attempts or carrying out attacks. As part of its automated security response, PfSense promptly blocks the IP addresses associated with these attacks. Both the log alerts and the blocked logs on Snort are available for download, facilitating network forensic analysis and aiding in the investigation of potential security incidents[31].

Each attack or attempted assault originating from a PC client has a discernible impact on traffic managed by PfSense, underscoring the importance of proactive security measures in maintaining the integrity and stability of the network infrastructure[32].

Fig.13. Attack Traffic

The attempted attacks originating from the client are observed on the LAN interface, indicating that the activities are occurring within the school network. The LAN graph clearly illustrates the traffic, reaching a peak of 12 Mbps, suggesting a significant level of activity on the interface. Furthermore, the traffic exhibits a noticeable surge, eventually stabilizing at a constant 12 Mbps. In contrast, normal network traffic, without any attack attempt, does not reach such high levels of 12 Mbps [33]. The alerts generated by Snort serve as crucial evidence for network forensic investigations[33]. Through this information, network administrators can gain insights into the ongoing events within the computer network. They can trace and analyze the data related to the attack and attempted attacks [34], enabling them to better understand the nature and source of the malicious activities. Armed with this knowledge, administrators can take appropriate measures to safeguard the network's security and prevent future incidents [35].

## V. CONCLUSION

The implementation of a forensic server using PfSense and Snort empowers network security by offering comprehensive attack prevention and investigation capabilities. Leveraging a built-in package manager, the system can efficiently detect and handle potential attacks. By selectively enabling rules tailored to the specific requirements, Snort becomes adept at identifying various attack types. Upon detection of an attack, PfSense promptly responds with automated preventive measures, such as blocking the offending traffic. Meanwhile, the alerts generated by Snort serve as valuable resources for network forensic investigations. These alerts provide crucial insights into the nature and scope of the attacks, enabling network administrators to conduct in-depth analyses and take appropriate actions to mitigate risks effectively. The combination of PfSense and Snort not only strengthens network security but also enhances the organization's ability to preemptively safeguard its assets from potential threats, making it a reliable and indispensable solution for protecting and maintaining a secure network environment.